\documentclass[twocolumn,showpacs,preprintnumbers,amsmath,amssymb]{revtex4}
\usepackage[english]{babel}
\usepackage[latin2]{inputenc}
\usepackage{epsfig}

\begin{document}

\title{Interstitial Mn in (Ga,Mn)As: Binding energy and exchange coupling}
\author{J.~Ma\v{s}ek and F.~M\'{a}ca}

\affiliation{Institute of Physics ASCR, Na Slovance 2, 182 21
Praha 8, Czech Republic}

\date{August 26, 2003}

\begin{abstract}
We present {\sl ab initio} calculations of total energies of Mn
atoms in various interstitial positions. The calculations are
performed by the full-potential linearized plane-wave method. The
minimum energy is found for tetrahedral T(As$_{4}$) position, but
the energy of the T(Ga$_{4}$) site differs by only a few meV. The
T(Ga$_{4}$) position becomes preferable in the p-type materials.
In samples with one substitutional and one interstitial Mn the Mn
atoms tend to form close pair with antiparallel magnetic moments.
\\
We also use the spin-splitting of the valence band to estimate the
exchange coupling $J_{pd}$ for various positions of Mn. The
exchange parameter is the same for the substitutional and for the
T(As$_{4}$) position and it is somewhat smaller in the case of the
T(Ga$_{4}$) position.
\end{abstract}


\maketitle


\section{Introduction}

Diluted magnetic semiconductors, such as (Ga,Mn)As, are important
materials combining ferromagnetic behavior with a sensitivity to
doping characteristic for semiconductors. This leads to their
interesting physical properties and makes possible their
applications e.g. in spin-electronics \cite{Dietl00, Ohno03}.
Although (Ga,Mn)As has been extensively studied in the last years,
some aspects of the incorporation of Mn into the crystal lattice
still remain unclear. It was assumed that in the well defined
samples Mn simply substitutes for the host cation. Only recently
it was pointed out \cite{Masek01, Maca02} that the presence of
interstitial Mn may explain some peculiar properties of (Ga,Mn)As
such as the low doping efficiency of Mn acceptors
\cite{Beschoten99}. The main reason for considering the Mn atoms
on the interstitial positions was that they act as double donors
and partly compensate the Mn acceptors in the substitutional
positions.

At the same time, channelling Rutherford backscattering
experiments proved that a large fraction of Mn atoms indeed
occupies the interstitial positions (Mn$_{\rm int}$) in the
as-grown samples \cite{Yu02}. A correlation between the removal of
Mn interstitials and increase of the conductivity, the Curie
temperature, and saturation magnetization has been found.

In addition to the self-compensation effect, the interstitial Mn
atom also reduce the number of local moments that participate in
the ferromagnetic state. This was explained by pairing of Mn$_{\rm
int}$ with the Mn atoms in the substitutional positions (Mn$_{\rm
Ga}$) due to their Coulomb attraction \cite{Yu02}. At the bonding
distance, the antiferromagnetic superexchange within the pair is
assumed to outweigh the hole mediated ferromagnetic exchange. As a
result, the moments of the paired Mn atoms have opposite
directions and the pair as a whole has no magnetic moment.

The {\sl ab initio} studies of the interstitial Mn showed the
differences in the electronic structure of the Mn atom in the
interstitial and substitutional positions \cite{Maca02,Sanvito02}
and also a possible reaction path for the incorporation of Mn into
the GaAs lattice \cite{Erwin02}. It was also found that the
increase of the lattice constant of Ga$_{\rm 1-x}$Mn$_{\rm x}$As
with increasing content of Mn is partly due to the presence of Mn
in the interstitial positions \cite{Masek03}.

Recently, Blinowski and Kacman \cite{Blinowski02, Blinowski03}
investigated the coupling of the local moment on the interstitial
Mn to the spin of the carriers in the valence band. Using a
simplified tight-binding model they showed that the local moment
on Mn atom in the tetrahedral position with four Ga neighbors,
T(Ga$_{4}$), is effectively decoupled from the spin of the holes
in the valence band.

They claimed that the coupling is weak because, in addition to the
reduced number of the holes, also the coupling constant $J_{pd}$
is small for the interstitial Mn in the T(Ga$_{4}$) position. As a
result, the local moment is not subject to the ferromagnetic
coupling with the moments around it. This opens the way for the AF
exchange to be important in the Mn$_{\rm int}$-Mn$_{\rm Ga}$ pair,
as anticipated in Ref. \cite{Yu02}.

There are, however, several open questions concerning the magnetic
interactions of Mn in the interstitial positions. First of all,
the spin-polarized band structures \cite{Maca02} did not show any
indication of different values of $J_{pd}$ for substitutional and
interstitial Mn, at least for the T(As$_{4}$) positions. In
addition, the Mn$_{\rm int}$-Mn$_{\rm Ga}$ pair as a whole is
expected \cite{Blinowski02, Blinowski03} to have only a small
magnetic moment, but the exchange coupling with Mn$_{\rm Ga}$ --
being uncompensated by the contribution of Mn$_{\rm int}$ --
should strongly polarize the holes.

That is why we performed a more detailed study of the interstitial
Mn and its spin interactions. We use the density-functional,
full-potential linearized augmented plane wave calculations
(FPLAPW \cite{WIEN}) to obtain the electronic structure of
(Ga,Mn)As with Mn atoms in various crystallographic positions. The
calculated total energies are used to compare different positions
of the interstitial Mn and to estimate the strength of the
Mn$_{\rm Ga}$-Mn$_{\rm int}$ pair interactions. The splitting of
the valence band for the majority- and minority-spin electrons is
used to compare the corresponding values of the exchange
parameters $J_{pd}$.


\section{M\lowercase{n} in various interstitial positions}

We compare the total energies for three interstitial positions of
Mn in GaAs. There are two inequivalent tetrahedral positions in
the zinc-blende structure of GaAs, T(As$_{4}$) and T(Ga$_{4}$).
They are surrounded by four As and Ga atoms, respectively. The
(unrelaxed) distances of these nearest neighbors are equal to the
length $d_{1}$ of the Ga-As bond, i.e. to the Mn-As distance for
the substitutional Mn$_{\rm Ga}$. The local arrangement around the
interstitial Mn is, in contrast to Mn$_{\rm Ga}$, characterized by
another six close neighbors at the distance $d_{2} = \sqrt{9/8}
\cdot d_{1} \approx 1.155 d_{1}$. In the hexagonal position, the
interstitial Mn has three Ga and three As atoms at the same
distance $d_{hex} = \sqrt{11/12} \cdot d_{1} \approx 0.957 d_{1}$,
and no other close neighbors.

We represent GaAs with Mn$_{\rm int}$ by hexagonal
Ga$_{12}$As$_{12}$Mn supercells. The c-axis coincides with the
body diagonal of the conventional cubic unit cell. The symmetry of
the cell does not change if we shift the Mn interstitial along the
c-axis from T(As$_{4}$) to hexagonal and T(Ga$_{4}$) positions.
This makes possible to perform all calculations under the same
conditions.


\begin{table}[h]
\caption{Total energy, $E_{tot}$, and total spin, $S_{tot}$, of
the unit cell of Ga$_{12}$As$_{12}$Mn with various interstitial
positions of Mn. Ga$_{9}$Zn$_{3}$As$_{12}$Mn samples are used to
simulate p-type materials.}
 \begin{tabular}{|c|c|c|c|}
 \hline
 Sample & Mn position & $E_{tot}$ (eV) & $S_{tot}$ \\
 \hline
 Ga$_{12}$As$_{12}$Mn & T(As$_{4}$)   & ground state & 1.665 \\
 Ga$_{12}$As$_{12}$Mn & T(Ga$_{4}$)   & + 0.005 & 1.555 \\
 Ga$_{12}$As$_{12}$Mn & hex.   & + 0.522 & 1.519 \\
 \hline
 Ga$_{9}$Zn$_{3}$As$_{12}$Mn & T(As$_{4}$)   & + 0.063 & 2.331 \\
 Ga$_{9}$Zn$_{3}$As$_{12}$Mn & T(Ga$_{4}$)   & ground state & 2.176 \\
 Ga$_{9}$Zn$_{3}$As$_{12}$Mn & hex.   & + 1.320 & 2.136 \\
 \hline
 \end{tabular}
\end{table}

The results are summarized in Table I. Surprisingly, the binding
energy of the interstitial Mn does not depend much on its nearest
neighbors. The interstitial Mn has minimum energy in the
T(As$_{4}$) position. However, the difference of the total
energies obtained for Mn in T(Ga$_{4}$) and T(As$_{4}$) positions
is of order of a few meV and can be neglected in practice. This
means that, without intervention of other charged defects, the Mn
interstitials can be found with an almost equal probability in
either T(As$_{4}$) or T(Ga$_{4}$) position. The total energy
corresponding to the hexagonal interstitial position of Mn is
remarkably higher and represents a barrier $\approx$ 0.5 eV
separating the tetrahedral positions. Calculations with similar
results have been recently done by Boguslawski \cite{Edmonds03}.

The weak influence of the nearest neighbors on the interstitial Mn
in T(As$_{4}$) and T(Ga$_{4}$) positions can be observed also in
the densities of states in Fig. 1. The total DOS and the
distribution of the Mn d-states are almost identical. Also the DOS
of Ga$_{12}$As$_{12}$Mn with Mn in the hexagonal position is quite
similar and differs mainly by the overlap of the valence band for
the majority-spin electrons ($\uparrow$) with the conduction band
for the minority-spin ($\downarrow$) electrons. This tendency to
close the gap in the electron spectrum correlates well with the
increase of the total energy.


\begin{figure}
\begin{center}
\includegraphics[width=50mm,height=90mm,angle=270]{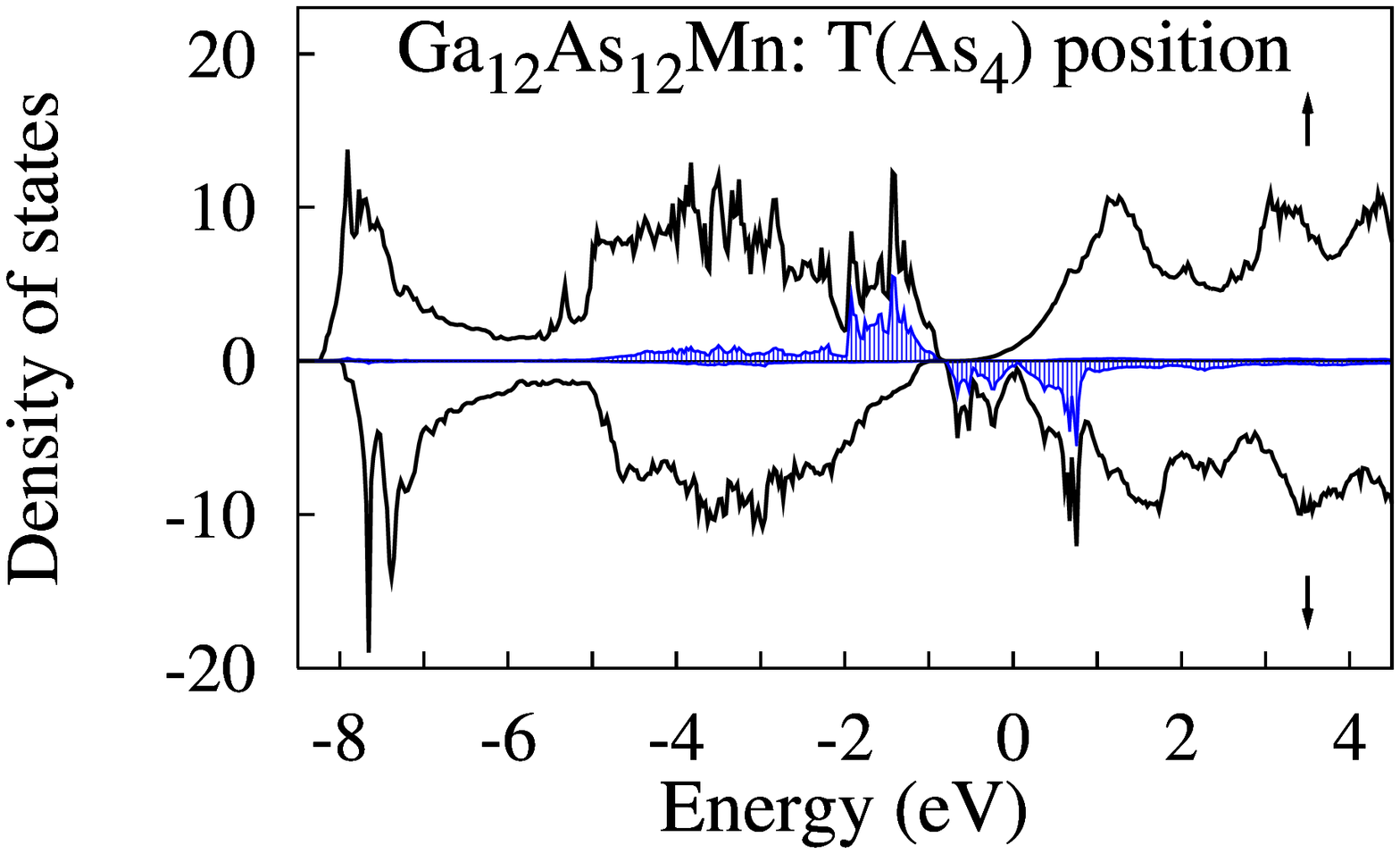}
\includegraphics[width=50mm,height=90mm,angle=270]{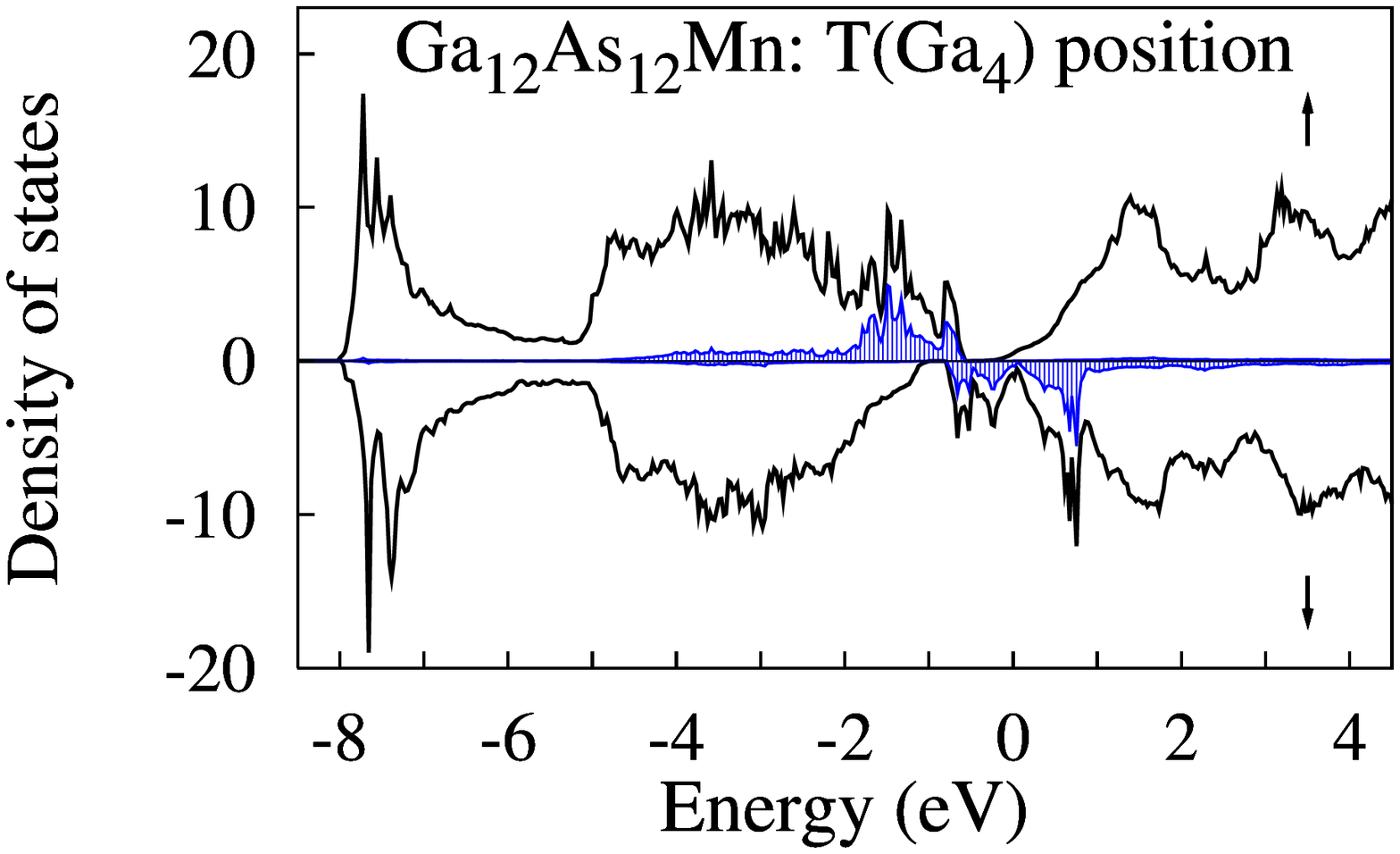}
\includegraphics[width=50mm,height=90mm,angle=270]{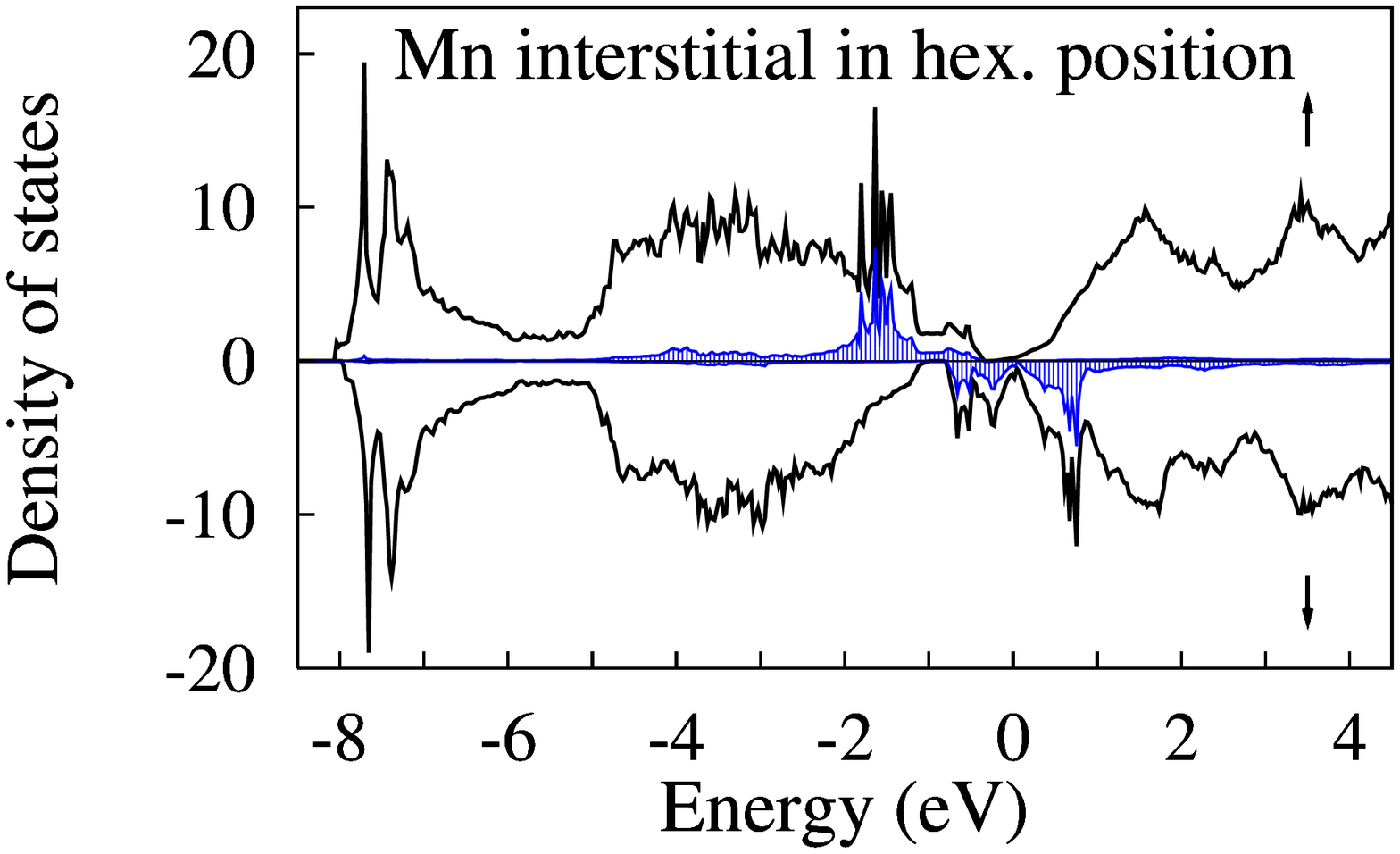}
\caption{Spin-polarized densities of states for
Ga$_{12}$As$_{12}$Mn with an interstitial Mn atom in T(As$_{4}$)
(upper panel), T(Ga$_{4}$) (middle panel), and hexagonal position
(lower panel). The contribution of Mn d-states is indicated by
hatched area.}
\end{center}
\end{figure}

The lattice relaxation around the Mn impurity is not very
important for the substitutional Mn as shown by Mirbt et al.
\cite{Mirbt02}. This is not the case of the interstitial Mn which
tends to expand the crystal lattice \cite{Masek03}. The addition
of Mn into either T(As$_{4}$) or T(Ga$_{4}$) positions results in
a remarkable repulsion of the nearest and next-nearest neighbors.
The positions of the more distant atoms in the supercell are
changed much less. The energy gain due to the relaxation is in
both cases approximately 20~meV and it does not change the
ordering  of the total energies for the T(As$_{4}$) or T(Ga$_{4}$)
positions. The lattice relaxation around Mn in the hexagonal
position has not been considered because of its minor importance.

In the case of the T(As$_{4}$) position, the distance of the four
nearest As neighbors was found to increase by $\approx$ 1.5~\%
from {2.45~\AA} to {2.47~\AA}. The distance of the six
next-nearest Ga neighbors increases by $\approx$ 0.7~\% from
{2.835~\AA}  to 2.88~\AA.

The relaxation around Mn in the T(Ga$_{4}$) position is larger.
The nearest Ga neighbors are pushed to the distance {2.515~\AA}
and the relaxed distance of the next-nearest neighbors increases
by $\approx$ 0.5~\%. The enhanced relaxation of the nearest
neighbors in this case is due to the Coulomb repulsion between Mn
and Ga atoms which are both positively charged.

In all cases, there are two electrons in the conduction band,
i.e., the Mn$_{\rm int}$ always acts as a double donor. The
electrons in the conduction band are almost completely
spin-polarized. They accumulate in the minority-spin conduction
band, so that the total spin of the cell (i.e., the spin per Mn)
is reduced to $\approx 3/2$ in accordance with our previous
calculations \cite{Maca02}.

For comparison, we performed the same set of calculations also for
hypothetical Ga$_{9}$Zn$_{3}$As$_{12}$Mn crystals with Zn atoms
substituted at the sites most distant from the interstitial Mn.
The presence of Zn has only a little effect on the density of
states, but the material is converted into the p-type with one
hole in the valence band. In this case, the T(Ga$_{4}$) turns to
be the stable interstitial position of Mn. The energy difference
between the T(As$_{4}$) and T(Ga$_{4}$) positions, approximately
60 meV,  is high enough for a preferential occupation of the
T(Ga$_{4}$) position at typical growth conditions. The remarkable
increase of the energy of the hexagonal position indicates that
the mobility of the interstitial Mn may depend on the type and
degree of the doping.


\section{Binding energy and exchange coupling of a M\lowercase{n} -- M\lowercase{n} pair}

The hexagonal unit cell used in Sect. II is well suited also for
the study of the Mn$_{\rm int}$-Mn$_{\rm Ga}$ pair interactions.
We consider a hypothetical Ga$_{11}$MnAs$_{12}$Mn crystal with one
substitutional and one interstitial Mn in the unit cell. We
consider three positions of the interstitial Mn shown in Fig. 2.
The positions T(Ga$_{3}$Mn) and T(As$_{4}$), denoted (a) and (b)
in Fig. 2, are located on the c-axis of our unit cell and
correspond to the initial stage of dissociation of the Mn$_{\rm
Ga}$-Mn$_{\rm int}$ pair. On the other hand, the T(As$_{4}$)
position denoted (c) is very close to Mn$_{\rm Ga}$ and is -
together with T(Ga$_{3}$Mn) - a candidate for the ground state of
Mn$_{\rm int}$.


\begin{figure}[b]
\begin{center}
\includegraphics[width=70mm,height=60mm,angle=0]{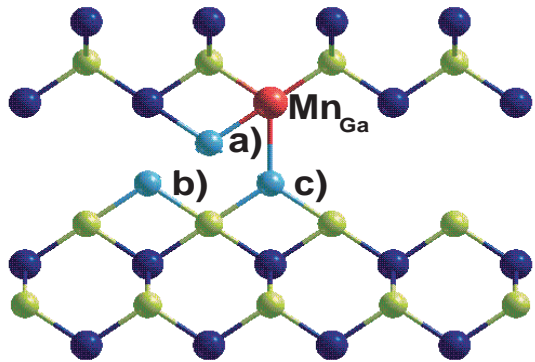}

\caption{Configurations of the Mn$_{\rm Ga}$-Mn$_{\rm
int}$ pair in the (110) plane of GaAs crystal \cite{XCrySDen99}:
(a) Mn$_{\rm int}$ in T(Ga$_{3}$Mn) position is the nearest
neighbor of Mn$_{\rm Ga}$; (b) Mn$_{\rm int}$ in T(As$_{4}$)
position representing a partially dissociated pair at a doubled
distance; (c)
 Mn$_{\rm int}$ in the T(As$_{4}$) position closest to Mn$_{\rm
 Ga}$.}
\end{center}
\end{figure}

In the T(Ga$_{3}$Mn) position, Mn$_{\rm int}$ and Mn$_{\rm Ga}$
are the nearest neighbors at the distance $d_{1}$. We performed
the density-functional calculations with two initial conditions,
corresponding to parallel and antiparallel alignment of their
local moments. In both cases, the self-consistent procedure
converges to a locally stable solution without changing the
initial alignment of the local moments. The resulting total
energies are given in Table II. The antiparallel alignment is
energetically more favorable than the parallel alignment, in a
good correspondence with the expectations \cite{Yu02,
Blinowski03}. The coupling is strong enough so that the AF state
of the Mn$_{\rm int}$-Mn$_{\rm Ga}$ pair is stable with respect to
the thermal fluctuations.

The Table II shows also the local spins on Mn$_{\rm Ga}$ and
Mn$_{\rm int}$ atoms defined as integrals of the spin density over
the corresponding atomic spheres. Although these quantities are
not directly related to the size of the observable local moments,
we can see that the local moments of Mn$_{\rm Ga}$ and Mn$_{\rm
int}$ are comparable and that the total magnetic moment of the
pair in the ground state is much smaller than the magnetic moment
of a single Mn.
\\


\begin{table}[h]
\caption{Total energy of Ga$_{11}$MnAs$_{12}$Mn unit cell and the
spin assigned to Mn$_{\rm int}$ and Mn$_{\rm Ga}$ atoms for
various configurations of the Mn$_{\rm int}$-Mn$_{\rm Ga}$ pair
shown in Fig 2. The configuration (c) is treated separately
because of its different symmetry.}

 \begin{tabular}{|c|c|c|c|c|}
 \hline
 Pair & Mn$_{{\rm Ga}}$ - Mn$_{{\rm int}}$ & Total energy
 & Spin & Spin \\
 config.     & distance (\AA) & (eV) & of Mn$_{\rm Ga}$ & of Mn$_{\rm int}$ \\
 \hline
 (a) $\uparrow \uparrow$ & 2.443 & + 0.324 & 1.678 & 1.617 \\
 (a) $\uparrow \downarrow$ & 2.443 & ground state & 1.778 & -1.531 \\
 (b) $\uparrow \uparrow$ & 4.886 & + 0.502 & 1.930 & 1.616 \\
 (b) $\uparrow \downarrow$ & 4.886 & + 0.330 & 1.899 & -1.558 \\
 \hline
 (c) $\uparrow \uparrow$ & 2.835 & + 0.303 & 1.774 & 1.657 \\
 (c) $\uparrow \downarrow$ & 2.835 & ground state & 1.842 & -1.549 \\
 \hline
 \end{tabular}
\end{table}

Assuming that the exchange coupling between more distant Mn atoms
from different unit cells is much smaller than the exchange
interaction $J(d_{1})$ within the closest pair we can estimate its
strength from the difference $\Delta E(d_{1}) = E_{\uparrow
\uparrow}(d_{1}) - E_{\uparrow \downarrow}(d_{1}) \approx -2
J(d_{1}) S^{2}$ of the total energies. Using saturated values $S =
5/2$ for both local moments we obtain the lower estimate for
$J(d_{1})$, namely $J(d_{1}) \approx$ -26~meV.

The coupling between Mn$_{\rm int}$ and Mn$_{\rm Ga}$ remains
antiferromagnetic also if Mn$_{\rm int}$ moves to any of the
adjacent T(As$_{4}$) positions (b) and (c) in Fig. 2. The energy
difference between the parallel and antiparallel alignment of the
local moments decreases with the increasing distance of the Mn
atoms, as expected for the superexchange. It remains almost the
same for the nearest and close next-nearest pairs and it is
reduced approximately to one half of $\Delta E(d_{1})$ for the
doubled distance corresponding to the configuration (b).

It should be pointed out, however, that the density-functional
calculations tend to overestimate the strength of the exchange
coupling. The reason for this is that the exchange splitting of
the Mn d-states, i.e. the separation of occupied majority-spin and
empty minority-spin states on the energy scale levels is
systematically underestimated. According to our calculations, the
exchange splitting $\varepsilon_{d}(\downarrow) -
\varepsilon_{d}(\uparrow)$ deduced from the spin-polarized
spectral distribution of Mn d-states  ranges from 2~eV to 3~eV.
This is roughly one half of the realistic estimate for the
exchange splitting in Mn (cf. e.g. Ref. \cite{Drchal91}).
Correspondingly, the above given value of $J(d_{1})$ should be
divided by four. In this way, we end with a value close to the
result of Ref. \cite{Blinowski03}.

Finally, we estimate the binding energy of the Mn$_{\rm
int}$-Mn$_{\rm Ga}$ pair. We compare the energies corresponding to
the (a) and (b) configurations from Fig. 2, both in the magnetic
ground state with the antiparallel alignment of the local moments.
The partially dissociated pair has a higher energy and the energy
difference $E_{\uparrow \downarrow}(d_{2}) -
E_{\uparrow\downarrow}(d_{1})$ is approximately 0.33 eV. It is,
however, only the lower estimate for the binding energy of the
Mn$_{\rm int}$-Mn$_{\rm Ga}$ pair because the dissociation of the
pair is far from being complete in our periodic model.
Nevertheless, even the value obtained here indicates that the Mn
interstitials are strongly attracted by the Mn$_{\rm Ga}$ atoms.
As long as the concentration of Mn$_{\rm int}$ is lower than the
concentration of Mn$_{\rm Ga}$ we can assume that most of Mn$_{\rm
int}$ atoms are involved in the pairs and that the blocking
mechanism proposed in Ref. \cite{Yu02} works.


\section{Exchange interaction of interstitial M\lowercase{n} and holes}

Due to the hybridization with the spin-polarized Mn d-states, also
the distribution of the states derived from the GaAs valence band
depends on the spin. In particular, the valence-band states for
the majority-spin electrons hybridize with the occupied d-states
and are pushed to higher energies. The minority-spin states are,
on the other hand, pushed down due to their hybridization with
unfilled d-states.
\\


\begin{table}
\caption{Spin splitting $\Delta E_{v} = E_{v}(\uparrow) -
E_{v}(\downarrow)$ for ferromagnetic state of (Ga,Mn)As.}
 \begin{tabular}{|c|c|c|c|}
 \hline
 Sample & Mn positions & $\Delta E_{v}$ & J$_{pd}$\\
 ~ & ~ & (eV) & (eVnm$^{3}$)\\
 \hline
 Ga$_{12}$MnAs$_{12}$ & T(As$_{4}$)   & 0.627 & 0.14 \\
 Ga$_{12}$MnAs$_{12}$ & T(Ga$_{4}$)   & 0.515 & 0.11 \\
 Ga$_{12}$MnAs$_{12}$ & hex.         & 0.653 & 0.14 \\
 \hline
 Ga$_{16}$MnAs$_{16}$ & T(As$_{4}$)   & 0.502 & 0.14 \\
 Ga$_{16}$MnAs$_{16}$ & T(Ga$_{4}$)   & 0.429 & 0.12 \\
 Ga$_{14}$Mn$_{2}$As$_{16}$ & 2 $\times$ Mn$_{\rm Ga}$ & 1.021 & 0.15 \\
 Ga$_{14}$Mn$_{3}$As$_{16}$ & 2 $\times$ Mn$_{\rm Ga}$ + T(As$_{4}$) & 1.328 & 0.13 \\
 Ga$_{30}$Mn$_{3}$As$_{32}$ & 2 $\times$ Mn$_{\rm Ga}$ + T(As$_{4}$) & 0.743 & 0.14 \\
 \hline
 \end{tabular}
\end{table}

This effect is formally described by the Kondo exchange
interaction between the local spins ${\bf S}_{i}$ at sites $R_{i}$
occupied by Mn and the spin density ${\bf s}(r)$ due to the
itinerant holes \cite{Moriya85},
\begin{equation}
H_{int} = J_{pd} \sum_{i} {\bf S}_{i} \cdot {\bf s}(R_{i}).
\end{equation}
The exchange parameter $J_{pd}$ characterizes the strength of the
coupling. Within the mean-field theory, the Kondo exchange
interaction results in the splitting of the valence band edge
$E_{v}$. The splitting $\Delta E_{v}$ is proportional to the size
$S$ of the local spins and to the concentration $x$ of magnetic
ions,
\begin{equation}\label{mfdev}
\Delta E_{v} \equiv E_{v}(\downarrow) - E_{v}(\uparrow) =
\frac{4x}{a^{3}} J_{pd} S
\end{equation}
assuming spin $\frac{1}{2}$ for the holes, $a$ is the lattice
constant. We use Eq. (\ref{mfdev}) to determine the exchange
parameter $J_{pd}$ from our spin-polarized band structures. The
results of the calculations for the unrelaxed geometries presented
here and in Ref. \cite{Maca02} are summarized in Table III.

The resulting values of $J_{pd}$ are overestimated from the same
reasons as discussed above in Sect. III. Being divided by a factor
of two (the reduction factor for the energy nominator), they
approach the realistic values \cite{Ohno99}. In this work,
however, we concentrate on the comparison of $J_{pd}$ for
different geometries. The overestimate of $J_{pd}$ due to the
reduced band gap is expected to be similar in all cases and not
very important in this respect.


\begin{figure}
\begin{center}
\includegraphics[width=52.5mm,height=81mm,angle=270]{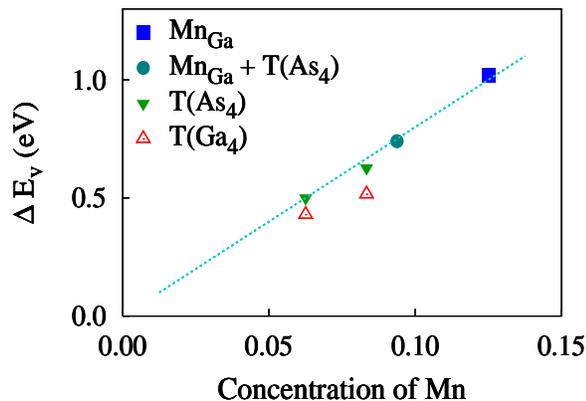}
\caption{Spin splitting of the valence band as a function of Mn
concentration.}
\end{center}
\end{figure}

In Fig. 3, we plot $\Delta E_{v}$ against $x$ to visualize the
linear increase of the band splitting with the concentration of
Mn. The most noticeable result is that all points in Fig. 3,
obtained for substitutional Mn as well for Mn in the interstitial
T(As$_{4}$) and T(Ga$_{4}$) positions lie around the same line.
This means that the value of $J_{pd}$ does not depend much on the
position of the Mn atom in the lattice.

This result is not very surprising in the case of the
substitutional Mn and T(As$_{4}$) position. In either case, Mn is
surrounded by four As atoms at the same distance $d_{1}$ and the
same degree of the hybridization of the valence band states with
Mn d-states can be expected.

The results for the T(Ga$_{4}$) position are not simple to
interpret. The value of $J_{pd}$ for both Ga$_{12}$MnAs$_{12}$ and
Ga$_{16}$MnAs$_{16}$ is smaller than but comparable with the
result for the substitutional Mn$_{\rm Ga}$ site. Assuming for
simplicity that only the hybridization of the Mn d-states with
p-states of the nearest neighbors is relevant for the spin
splitting $\Delta E_{v}$, the values of $J_{pd}$ simply reflect
the orbital composition of the valence band. It is well known that
the top of the valence band in GaAs, as well as in other III-V
semiconductors, is composed of both anion and cation p-states, in
the proportion approximately $c_{p}^{\rm As} \approx 3/4$,
$c_p^{\rm Ga} \approx 1/4$ \cite{Harrison80}. This means that the
same proportion (roughly 3:1) should be expected also for the
exchange parameters $J_{pd}$ parameters corresponding to Mn atoms
in substitutional and T(Ga$_{4}$) positions. The strong deviation
of the actual density-functional results from this model
expectation may indicate that the hybridization of the Mn d-states
with more distant neighbors has a remarkable influence on the
magnetic interactions.

The fact that the magnetic behavior of the interstitial Mn is
rather insensitive to its position in the crystal has two
important implications concerning the Mn$_{\rm Ga}$-Mn$_{\rm int}$
pairs. First of all, the pair as a whole does not interact much
with the spin of the holes because the effects due to Mn$_{\rm
Ga}$ and Mn$_{\rm int}$ compensate one another. In this respect,
our {\sl ab initio} results overcome the troubles of the
simplified tight-binding models mentioned in the Introduction.

In addition, the effective 'annihilation' of Mn$_{\rm Ga}$ due to
pairing with Mn$_{\rm int}$ is not restricted to the closest pairs
with Mn$_{\rm int}$ in the T(Ga$_{3}$Mn) position, but it works as
long as the Mn atoms are close one to another and the exchange
coupling in the pair remains antiferromagnetic.


\section{Conclusions}

We used the FPLAPW method to obtain total energies of supercells
simulating various geometric and magnetic configurations of Mn
atoms in (Ga,Mn)As. In absence of other defects, the ground state
of the Mn interstitials is the tetrahedral T(As$_{4}$) position.
The energy of the T(Ga$_{4}$) position, however, is almost the
same. The situation changes in the p-type material where the
T(Ga$_{4}$) position has a lower energy. The hexagonal
interstitial position has much higher energy and represents a
barrier for diffusion of Mn from one to another interstitial site.
The barrier, and consequently also the mobility of the
interstitial Mn, depends on the doping.

The exchange coupling $J_{pd}$ of Mn interstitials with the holes
in the valence band is, for both T(As$_{4}$) and T(Ga$_{4}$),
close to the value of $J_{pd}$ obtained for the substitutional Mn.
This is not consistent with the simplest tight-binding picture of
Mn d-states that hybridize only with the nearest neighbors. In
this way, our result indicate that the hybridization with more
distant neighbors may be also important for the magnetic
interactions.

The Mn interstitials are attracted to the substitutional Mn and
form stable and magnetically inactive pairs. The
density-functional estimate for both binding energy of the pair
and for the energy of the antiferromagnetic coupling is of order
of 0.3 eV. This fits well with the present day notion of the
interstitial Mn in (Ga,Mn)As. \cite{Yu02, Blinowski03}. In
contrary to the general opinion, however, we found that the
efficient pairing is not restricted to Mn$_{\rm int}$ in the
T(Ga$_{3}$Mn) position and we showed the importance of the close
next-nearest neighbors for the properties of the interstitial Mn.

\subsection*{Acknowledgement}

This work has been done within the project AVOZ1-010-914 of the
ASCR. The financial support provided by the Grant Agency of the
ASCR (Grant No. A1010214) and by RTN project No.
HPRN-CT-2000-00143 the EC is acknowledged.



\end{document}